\documentclass[aps,prl,showpacs,showkeys,twocolumn,groupedaddress]{revtex4}

\usepackage{amsmath}
\usepackage{amsfonts}
\usepackage{amssymb}
\usepackage{latexsym}
\usepackage{graphicx}
\usepackage{float}
\usepackage{wrapfig}

\newcommand{\eins}{\leavevmode\hbox{\small1\kern-3.8pt\normalsize1}}
\newcommand{\R}{\scriptstyle{\mathrm{R}}}

\begin{document}

\title{Collective versus Single--Particle Motion in Quantum Many--Body
  Systems: Spreading and its Semiclassical Interpretation} 
\author{Jens H\"ammerling, Boris Gutkin and Thomas Guhr} 
\date{\today}
\affiliation{Fakult\"at f\"ur Physik, Universit\"at Duisburg--Essen,
                    Lotharstra\ss e 1, 47048 Duisburg, Germany}

\begin{abstract}
  We study the interplay between collective and incoherent
  single--particle motion in a model of two chains of particles whose
  interaction comprises a non--integrable part. In the perturbative
  regime, but for a general form of the interaction, we calculate the spectral density for collective
  excitations.  We obtain the remarkable result that it always has a
  unique semiclassical interpretation. We show this by a proper
  renormalization procedure which allows us to map our system to a
  Caldeira--Leggett--type of model in which the bath is part of the
  system.
\end{abstract}
% 05.45.Tp   Time series analysis 
% 02.50.-r     Probability theory, stochastic processes, and statistics 
% 02.50.Sk   Multivariate analysis
% 02.50.Tt    Inference methods 
% 02.20.-a    Group theory 
% 02.20.Qs   General properties, structure, and representation of Lie groups 

\maketitle

Collective motion, \emph{i.e.}, coherent motion of the particles in
phase space, is a fundamental feature of many--body systems.  A wealth
of information on collectivity is available for atomic nuclei
\cite{Bohr}.  Bose--Einstein condensates \cite{But99,Mad00,Mar00}
provide more recent examples.  There is strong experimental
\cite{Oos99} and theoretical \cite{Tor04} evidence that similar
effects occur in other fermionic systems as well.  Collective motion
emerges out of the incoherent, single--particle motion whenever
favored by energy and kinematic conditions.  Due to the
quantum--classical correspondence principle, the collective dynamics on
the classical level should be reflected in the spectral properties of
the corresponding quantum many--body system.  Hence, the spectrum of a
many--body system comprises states of single--particle and of
collective character, mixed forms with a partial degree of
collectivity exist as well.  The details strongly depend on how the
system is probed.

Consider the Giant Dipole Resonance in heavier nuclei as a prominent
example which also serves as an inspiration for our model to be
discussed in the sequel. The cross section of electric dipole
radiation and the spectral density of the excitations show at a
certain energy a huge peak whose spreading width is orders of
magnitudes larger than the mean level spacing. It can be understood
in terms of the following picture: the neutrons are confined to one
sphere, the protons to another one. There is no or very little
relative motion of the nucleons inside these spheres. The two spheres,
however, move against each other, resulting in an enormous response
function. The difference between the center--of--mass coordinates of
the two spheres is the proper collective coordinate.  Many other forms
of collective motion in nuclei exist.

Not surprisingly, it is a demanding challenge to understand the
emergence of collective motion and its interplay with the incoherent
single--particle motion.  The vast majority of studies in this context
relies on effective models whose justification is often mainly
phenomenological or even on the level of hand waving if the system in
question is too complex. Better understanding of these issues is
called for.  In the present contribution, we have three goals: (i) We
want to address the interplay between collective and single--particle
motion from first principles in the framework of a tractable, yet
sufficiently general and complex model.  (ii) We aim at doing this
analytically in such a way that we identify the collective coordinate,
but always keep full control over the single--particle degrees of
freedom. (iii) We wish to deliver the important insight that the
spectral density of the collective excitations is directly related
to classical motion.

We begin with setting up our model which considerably generalizes an
integrable model which we studied previously~\cite{hgg}. In one
dimension, two chains $a=1,2$ of $N$ interacting particles each with
positions $x_i^{(a)}, \ i=1,\ldots,N$ and momenta $p_i^{(a)}, \
i=1,\ldots,N$ are coupled to one another. The total Hamiltonian reads
\begin{equation}
H = H_0 + \lambda  H_{1} \ .
\label{Hamiltonian}
\end{equation}
Here, $H_0= H_{0}^{(1)}+ H_{0}^{(2)} + H_{0}^{(12)}$ is the integrable
part considered in Ref.~\cite{hgg}.  The first two terms
\begin{equation}
H_{0}^{(a)} = \frac{1}{2m}\sum^N_{i=1}\left(p^{(a)}_i\right)^2 +
                                              \sum^N_{i,j=1}x^{(a)}_i W_{ij} x^{(a)}_j 
\label{eq2}
\end{equation}
model the two chains $a=1,2$ before they are coupled. The interaction
within each chain is harmonic and described by the matrix $W$ which we
assume equal in both chains. We are mainly interested in selfbound
systems such as nuclei, where unlike Bose--Einstein condensates no
external confining potential is needed. It is easy to impose corresponding conditions
on the matrix $W$ which ensure that the interaction is invariant
under translations and the system is bounded~\cite{hgg}.  We now couple the two chains by
an interaction which depends on the differences between their coordinates, 
\begin{equation}
H_{0}^{(12)} = \sum^N_{i,j=1}K_{ij} \left(x^{(1)}_i - x^{(2)}_j \right)^2  \ .
\label{eq3}
\end{equation}
For every choice of the coupling matrix $K$, the Hamiltonian $H_0$
is translation invariant. Clearly, the model is up to now integrable.

We generalize the model by adding the translation invariant term
$\lambda H_{1}$ which breaks integrability,
\begin{equation}
H_{1} = \sum^N_{i,j=1} f(x^{(1)}_i -x^{(2)}_j) \ ,
\label{eq4}
\end{equation}
where $f$ is an arbitrary, positive and even analytical function of the form
\begin{equation} 
f(z) = \sum^{\infty}_{n=2} f_n z^{2n} \ . 
\label{f_expansion}
\end{equation}
We introduce the parameter $\lambda$, because we aim at a
perturbative discussion.

To quantize our model, we replace coordinates and momenta by operators
$\hat{x}_i^{(a)}$ and $\hat{p}_i^{(a)}$. Importantly, we make this
step on the level of the \emph{original} particle degrees of freedom.
Motivated by the above mentioned Giant Dipole Resonance, we aim at
studying collective excitations and the associated spectral density.
We expect that it shows a pronounced peak, which we wish to understand
in (semi)classical terms.  Naturally, the collective coordinate $X$ is
the difference between the mass centers of the two chains, and it is
convenient to rescale it with a factor $\sqrt{N/2}$,
\begin{eqnarray}
X = \frac{1}{\sqrt{2N}}\left( \sum_{i=1}^{N} x^{(1)}_{i} -   
              \sum_{i=1}^{N} x^{(2)}_{i}\right) 
\label{eq5}
\end{eqnarray}
and accordingly for the collective operator $\hat{X}$.  To probe the
existence of quantum collective states in the excitation spectrum, we
investigate the correlator
\begin{equation}
S(t) = \langle \Phi_0|\hat{X}(t)\hat{X}(0)|\Phi_0\rangle
\label{eq6}\ ,
\end{equation}
where $|\Phi_0\rangle$ is the ground state of the total Hamiltonian
$\hat{H}$.  Here, $\hat{X}(t)$ is the Heisenberg picture of the
operator $\hat{X}$ with the time evolution governed by the total
Hamiltonian $\hat{H}$.  The Fourier transform of the correlator
\eqref{eq6},
\begin{equation}
\tilde{S}(\omega) = \sum_{\mu=0}^\infty|\langle \Phi_0|\hat{X}|\Phi_\mu\rangle|^2
               \delta\left(\omega-\frac{E_{\mu}-E_{0}}{\hbar} \right) \ , 
\label{FourierS}
\end{equation}
is the desired spectral density of the collective excitations. It
measures the strength of the transitions between the ground and the
excited states $|\Phi_\mu\rangle$ of the whole system and can be
interpreted as the response of the system that is excited by the
transition operator $\hat{X}$.  Following the terminology in
many--body physics, we say that there is a collective quantum state
for an energy $E_{{\rm col}}=E_{0}+\hbar\omega$ if $\tilde{S}(\omega)$
(smoothened over some energy interval) has a pronounced spike at the
corresponding frequency $\omega$.
  
To leading order $\lambda$ in perturbation theory, we obtain the
following expression for the correlator
\begin{multline}
 S(t) \approx \langle 0| \hat{X}_I(t) \hat{X}_I(0) |0\rangle \\ 
  + \lambda \sum_{l\neq 0} \left(a^1_{l0} \langle 0|  \hat{X}_I(t) \hat{X}_I(0) |l\rangle
 + (a^1_{l0})^* \langle l|  \hat{X}_I(t) \hat{X}_I(0) |0\rangle \right)\\
%  + 2 \lambda \mathrm{Re} \sum_{l=1}^\infty a^1_{l0} \langle 0|  \hat{X}_I(t) \hat{X}_I(0) |l\rangle\\
  + \frac{i\lambda}{\hbar}\int\limits_0^t dt_1 
        \langle 0| [ H_I(t_1), \hat{X}_I(t)] \hat{X}_I(0) |0\rangle + {\cal O}(\lambda^2) \ ,
\label{Perturbation}
\end{multline}
where the sum runs over the eigenstates $|l\rangle$ of $\hat{H}_0$,
and where $|0\rangle$ is the ground state of $\hat{H}_0$.  The
coefficients $a^1_{l0} = \langle l|\hat{H}_1 |0\rangle/(E_0 - E_l) $
turn out to be real due to time reversal invariance.  The sum in Eq.~\eqref{Perturbation} arises from the correction to the
ground state while the last term results from the perturbation of the
Hamiltonian.  The collective operator $\hat{X}_I$ and the
non--integrable part $\hat{H}_{1I}$ of the Hamiltonian appear in the
interaction picture whose time evolution is governed by the integrable
Hamiltonian $\hat{H}_0$. For any operator $\hat{F}$, we have
\begin{equation}
\hat{F}_I(t) = \exp\left(\frac{i}{\hbar}\hat{H}_0 t\right) \hat{F} 
                                     \exp \left(-\frac{i}{\hbar}\hat{H}_0 t\right) \ .
\label{eq6a}
\end{equation}
For later purposes, it is useful to consider the imaginary part of the
correlator, $\mathrm{Im} S(t) = S_1(t)$. We notice that, by
virtue of Eq.~\eqref{FourierS}, the Fourier transforms of $S(t)$ and
$S_1(t)$ are connected through
\begin{equation}
\tilde{S}(\omega) = 2i\Theta(\omega)\tilde{S}_1(\omega) \ , 
\end{equation}
where $\Theta(\omega)$ denotes the Heaviside step function.  Since
$\mathrm{Im} \langle 0| \hat{X}_I(t) \hat{X}_I(0) |l\rangle$ vanishes
for $l\neq 0$, the imaginary part of the correlator simplifies, and we
find
\begin{multline}
S_1(t) \approx \mathrm{Im} \langle 0| \hat{X}_I(t) \hat{X}_I(0) |0\rangle  \\
+\frac{\lambda}{2\hbar} \int\limits_0^t dt_1
      \langle 0|\left [ [ \hat{H}_{1I}(t_1), \hat{X}_I(t)], \hat{X}_I(0) \right] |0\rangle  + {\cal O}(\lambda^2) \ . 
\label{S_one}
\end{multline}
At this point, we may use our previous results \cite{hgg}. The first
term of Eq.~\eqref{S_one} was evaluated by mapping $H_0$ into the form
of a Caldeira--Leggett--like model, in which $X$ can be viewed as the
coordinate of a ``big'' particle in a harmonic potential which is coupled
to a ``bath'' of harmonic oscillators. The interpretation of the
``bath'', however, differs significantly form the standard
Caldeira--Leggett situation \cite{Caldeira:1982iu}.  In our case, the ``bath'' is not
external, it is part of the system and formed by the internal degrees of freedom.  The resulting expression for the
spectral function $\tilde{S}(\omega)$ is then
\begin{equation}
\tilde{S}(\omega) \approx \frac{\hbar}{\pi m} \Theta(\omega)\ 
           \mathrm{Im}\, \frac{1}{\Omega_0^2 - \omega^2 -
                          i\omega \tilde{\gamma}(\omega) } \ ,
\label{S_omega}
\end{equation}
where $\tilde{\gamma}$ formally coincides with the classical
``damping'' kernel, but here it describes the spreading of the
collective excitation over the spectrum. There is not an energy loss
or any kind of dissipation in our system.  The resonance frequency
$\bar{\Omega}_0$ is the fundamental oscillator frequency of the
corresponding classical problem.  To illustrate this result, we
mention that, when the collective degree of freedom $X$ interacts with
an ohmic ``bath'' (see Ref.~\cite{Breuer}), the spreading
kernel $\tilde{\gamma}(\omega)= \gamma_0$ is a constant and
$\tilde{S}(\omega)$ has a Lorentzian shape with the width $\gamma_0$
at the position of the classical oscillator frequency $\bar{\Omega}_0 = \sqrt{\Omega^2_0 - (\gamma_0/2)^2}$.

We now consider the crucial second term on the right hand side of
Eq.~\eqref{S_one}.  A naive continuation of the approach in
Ref.~\cite{hgg} quickly becomes cumbersome. Luckily, there is much
better way of tackling this term which eventually leads to a 
new insight to our problem. The second term is easily seen to be of
the form $\frac{\lambda}{2\hbar}\int_0^t dt_1 \chi(t_1,t),$
%\begin{equation}
%\frac{\lambda}{2\hbar}\int\limits_0^t dt_1 \chi(t_1,t),
%\label{eq7}
%\end{equation}
with the kernel $\chi(t_1,t)$ given by
\begin{equation}
\sum_{i\ne j} \langle 0 | \left[ [ f(\hat{x}^{(1)}_{i}(t_1) - 
                   \hat{x}^{(2)}_{j}(t_1)), \hat{X}(t)], \hat{X}(0) \right]|0\rangle \ . 
\label{R_Kernel}
\end{equation}
Owing to the harmonic form of $\hat{H}_0$, we find the identity
\begin{equation}
[(\hat{x}^{(1)}_{i} - \hat{x}^{(2)}_{j})^n, \hat{X}(-t)]  = 
n (\hat{x}^{(1)}_{i} - \hat{x}^{(2)}_{j})^{n-1}\beta_{ij}(t)
\label{eq8}
\end{equation}
for the commutator of the $n$--th powers of differences with the
collective operator.  For all $n$, it can be reduced to
the operator $(\hat{x}^{(1)}_{i} - \hat{x}^{(2)}_{j})^{n-1}$ 
multiplying the function $\beta_{ij}(t)$ defined by
$[(\hat{x}^{(1)}_{i} - \hat{x}^{(2)}_{j}), \hat{X}(-t)]=\beta_{ij}(t) \mathbf{1}$.
We notice that the commutator in the latter expression is proportional
to the unit operator $\mathbf{1}$.  Applying this formula twice yields
\begin{equation}
\chi(t_1,t)= \beta_{ij}(t_1-t) \beta_{ij}(t_1) C_{ij} \ ,
\label{eq9}
\end{equation}
where the elements of the matrix  $C$ are given by
the ground state expectation values involving the second derivative
of the arbitrary function $f$ defining the non--integrable
part of the interaction in Eq.~\eqref{eq4},
\begin{eqnarray}
C_{ij} = \langle 0| f''(\hat{x}^{(1)}_{i} - \hat{x}^{(2)}_{j}) |0\rangle \ . 
\label{C_matrix}
\end{eqnarray}
We emphasize that this result is not due to an expansion of the
function $f(z)$, it applies in leading order $\lambda$ to all functions
of the form \eqref{f_expansion}.

We arrive at the important insight anticipated above: precisely the
same equation for the kernel $\chi(t_1,t)$ follows when using the
harmonic Hamiltonian
\begin{eqnarray}
\hat{H}_0^{\mathrm{R}} = \hat{H}_{0} + 
                   \frac{\lambda}{2} \sum_{i,j=1}^N C_{ij} (\hat{x}^{(1)}_{i} -\hat{x}^{(2)}_{j})^2 \ .
\end{eqnarray}
In other words, the effect of a general, non--integrable perturbation
can, to leading order $\lambda$, be fully accounted for by a proper
renormalization of the integrable Hamiltonian $\hat{H}_0$. Since
$\hat{H}_0^{\mathrm{R}}$ is harmonic, the spectral function
$\tilde{S}(\omega)$ is given by Eq.~\eqref{S_omega}, where the
renormalized oscillation frequency $\Omega_0^{\mathrm{R}}$ and the
spreading kernel $\gamma^{\mathrm{R}}$ depend on the constant matrix
elements $C_{ij}$.  

The $C_{ij}$ themselves depend on the parameters of the integrable
Hamiltonian $\hat{H}_0$. Starting from the definition
\eqref{C_matrix}, we derive
\begin{eqnarray}
C_{ij} = \frac{1}{\sqrt{2\pi}} \int\limits_{-\infty}^{+\infty} 
            f''(x\sqrt{\alpha_{ij}}) \exp\left(-\frac{x^2}{2}\right) dx \ ,
\label{C_coeff}
\end{eqnarray}
where the quantities $\alpha_{ij}=\langle 0|{(x_{i}^{(1)} -
  x_{j}^{(2)})^2}|0 \rangle$ can be related to the parameters of
$\hat{H}_0$. They are given by
$\alpha_{ij}=(\Gamma_{ii}+\Gamma_{jj})/4$, where $\Gamma_{kk}$ is the
$k$-th diagonal element of the matrix
\begin{equation}
  \Gamma=\frac{\hbar}{\sqrt{2m}} 
        \left( \left( \frac{1}{W+M-K}\right)^{1/2}+
                      \left(\frac{1}{W+M+K}\right)^{1/2}\right) \ , 
\label{EqForGamma}
\end{equation}
and $M$ is a diagonal matrix whose elements read \cite{hgg} $M_{ij}=\delta_{ij}\sum_{l}{K}_{il}$.
The coefficients $C_{ij}$ depend on $\hbar$ since they result from
sandwiching the function $f$ with the ground states of the integrable
Hamiltonian. If we restrict ourselves to leading order $\hbar$, only
the first term of the expansion \eqref{f_expansion} enters.  In such
a semiclassical regime Eq.~\eqref{C_coeff} simplifies and we have $C_{ij}=3f_2(\Gamma_{ii}+\Gamma_{jj})$.

Since $H_0^{\mathrm{R}}$ is harmonic, the spectral function
$\tilde{S}(\omega)$ can now be calculated using Eq.~\eqref{S_omega}.
The renormalized oscillator frequency $\Omega_0^{\mathrm{R}}$ and
the renormalized spreading kernel $\gamma^{\mathrm{R}}$ are determined by the
renormalized coupling constants $K^{\R}_{ij}=K_{ij}+\lambda C_{ij} /2$
and by $W_{ij}$. The spreading kernel can be expressed
through the spectral density function $\sigma^{\R}(\omega)$ \cite{Breuer},
\begin{equation}
  \gamma^{\mathrm{R}}(t) =\frac{2}{m} \int\limits_0^{\infty}
        \frac{ \sigma^{\R}(\omega)}{\omega}\cos(\omega t) d\omega \ .
\label{damping_kernel}
\end{equation}
Employing our previous results~\cite{hgg}, we express the spectral
density function through the parameters of the harmonic Hamiltonian.
This yields
\begin{eqnarray}
\sigma^{\R}(\omega) =
-\frac{1}{2\pi m\omega} \mathrm{Im}\, \mathbf{k}^T 
         \frac{\eins}{(\omega+ i \epsilon)\eins-  (2 K_r/m)^{1/2}}\mathbf{k}, 
\label{densityexpression}
\end{eqnarray}
where ${K}_r$ and $\mathbf{k}$ are obtained from the matrix 
$
\tilde{K}^{\R}=A^T(W+M^{\R} +K^{\R})A$,
with $A$ being a discrete $N\times N$ cosine transform (DCT), see Ref.~\cite{hgg}. More precisely, ${K}_r$ is the $(N-1)\times(N-1)$ matrix
obtained from $\tilde{K}^{\R}$ by deleting the first row and the first
column while the vector $\mathbf{k}$ is the first column (excluding
the first element) of $\tilde{K}^{\R}$.  The renormalized oscillator
frequency turns out to be
\begin{equation}
 \Omega^{\R}_0=\sqrt{2\tilde{K}^{\R}_{11}/m +\gamma^{\R}(0)} \ . 
\label{renfrequency}
\end{equation}
It is important to notice that the spreading kernel
\eqref{damping_kernel}, the oscillator frequency \eqref{renfrequency}
and therefore the spectral density $\tilde{S}(\omega)$ are fully determined by the \emph{classical} dynamics of the renormalized
Hamiltonian $\hat{H}_0^{\mathrm{R}}$. In particular, using our
previous results \cite{hgg},  the  equation for
the classical time evolution  of the collective coordinate  reads
\begin{equation}
\frac{d^2 {X}(t)}{dt^2} + (\Omega^{\R}_0)^2 {X}(t) + 
\int\limits_0^t \gamma^{\R}(t-s) \frac{d{X}(s)}{ds} ds = 0 \ .
\label{classical}
\end{equation}
The same equation holds for the expectation value $\langle\hat{X}\rangle(t)$ if the initial state of the system  is properly chosen.
Equation~\eqref{classical} describes a damped, \emph{i.e.} in our case spread, harmonic
oscillator whose spreading kernel $\gamma^{\R}(t)$ and oscillator
frequency $\Omega^{\R}_0$ are given by Eqs.~\eqref{damping_kernel} and
\eqref{renfrequency}, respectively. Importantly, the solution of Eq.~(\ref{classical}) also determines, for properly chosen initial conditions, the spectral density of the collective excitations.
 We notice, however, that
$\hat{H}_0^{\mathrm{R}}$ itself contains quantum corrections which
depend on $\hbar$. Put differently, $\hat{H}_0^{\mathrm{R}}$ is
identified as the proper effective Hamiltonian whose classical
dynamics --- rather than the classical dynamics of the original, total
Hamiltonian $\hat{H}$ --- determines the spectrum of collective
excitations in leading order $\lambda$.
We also notice that higher order terms in the perturbative treatment of $S(t)$ come with higher
powers of $\hbar$. Indeed, it is straightforward to see that in the
case of $f(z)=f_2 z^4$ the $n$--th term of the perturbative expansion
scales as $(\hbar\lambda)^n$.  In this sense the renormalized
Hamiltonian $\hat{H}_0^{\mathrm{R}}$ provides the first semiclassical
correction to the spectrum of the collective modes.

We briefly discuss the conditions for the validity of our perturbative
approach.  The approximation (\ref{Perturbation}) can be used if the
following conditions are satisfied, see Ref.~\cite{Landau}.
First, the gap between the ground state and the first excited state of
$\hat{H}_0$ must be sufficiently large, i.e., $\lambda |\langle
0|\hat{H}_1|0\rangle|\ll\hbar\omega_{\min}$, where $\omega_{\min}$ is
the minimal oscillator frequency of the classical system given by the
lowest eigenvalue of the matrix $W+M+K$.  Second, the time $t$ of
propagation must be bounded by $\lambda |\langle
0|\hat{H}_1|0\rangle|t/\hbar\ll1$. As we are interested in time scales
of order $t\sim \Omega_0^{-1}$, the first condition implies the second
one.  Under these conditions the spectral characteristics such as
energy and spreading width of the collective excitations are close to
their values for the unperturbed Hamiltonian $\hat{H}_0$. We notice, however, that such a ``small'' perturbation $\lambda H_1$ might be quite large on the scale of the mean level spacing which is of the
order $\hbar^N$. This means that the local distribution of the energy
levels of the Hamiltonian $H$ might be essentially different from the
energy level distribution of the integrable Hamiltonian $H_0$.

%{\bf shouldn't that be our next paper??? - we should leave that out ..}
%However, such a ``small'' perturbation
%$\lambda \hat{H}_1$ might in fact be quite large on the scale of the mean level
%spacing which is of the order $\hbar^N$. This means that the local
%distribution of the energy levels of the Hamiltonian $H$ might be
%essentially different from the energy level distribution of the
%integrable Hamiltonian $\hat{H}_0$.
%{\bf .. up to here}

In conclusion, we studied, in the framework of a simple model, the
emergence of collectivity from first principles. We did not start from
an effective model, we rather derived an effective description and
still kept full control over the original degrees of freedom. In doing
so, we relate the expectation value of the collective operator and the spectral density of the collective excitations to a purely \emph{classical} equation.  We consider
that to be important, as it can be viewed as a justification of the
routinely used strategy in many--body physics, where effective models
are set up classically and then quantized.

Beside the fundamental aspects just mentioned, there is further
motivation for our study: Statistical analysis of the spectra
indicates that collective motion is typically regular while the
incoherent single--particle motion yields spectral statistics
described by random matrices, see Refs.~\cite{GMGW,End, End2}.  This
coexistence of both regular and chaotic dynamics in the same system is
a truly intriguing dynamical aspect of many--body
systems~\cite{Brack}. The regularity of collective motion implies that
the recent arguments \cite{hmabh} strongly supporting the
Bohigas--Giannoni--Schmit conjecture for single--particle systems do
not carry over in a straightforward manner to many--body systems. In short, this conjecture states that the spectral statistics of the single--particle system is of random--matrix type if the corresponding classical system is chaotic. Our study is thus needed when addressing the role of collectivity in quantum chaos.

We acknowledge support from Deutsche Forschungsgemeinschaft
(Sonderforschungsbereich Transregio 12 ``Symmetries and Universality
in Mesoscopic Systems'').

%%\bibliographystyle{unsrt}
%%\bibliography{Referenzen}

\end{document}